\documentclass[prb]{revtex4}
\usepackage{epsfig}

\begin{document}
\title{Waveguide containing  a backward-wave slab}
\author{I.S. Nefedov} 
\affiliation{Radio Laboratory, Helsinki University of Technology, P.O. 3000, FIN-02015 HUT,
Finland}
\author{S.A. Tretyakov}
\affiliation{Radio Laboratory, Helsinki University of Technology, P.O. 3000, FIN-02015 HUT,
Finland}
\email{sergei.tretyakov@hut.fi}

\begin{abstract}

We have considered theoretically the waveguide properties of a
plane two-layered waveguide, whose one layer is a usual magnetodielectric
(forward-wave medium), but another one is a slab of  so-called {\it backward-wave
material} (BW-material), whose both permittivity and permeability
are negative. We have analyzed the properties of eigenwaves in this
waveguide. In particular, it was found that
there exist  waves of both  TE and TM polarizations, whose fields decay exponentially from the interface
 of the two slabs inside both layers, and their slow-wave factor
 tends to infinity at small frequencies. Thus, this waveguiding system
 supports super-slow waves with extremely short wavelengthes,
 as compared to the free-space wavelength and the cross section size.
 Other peculiarities of the spectrum are also discussed.

\end{abstract}


\maketitle

\section{Introduction}

In recent years we have witnessed increased interest in electromagnetic
properties of
exotic media. In particular, media with  both
permittivity and permeability real and negative have been studied. Plane
electromagnetic waves in isotropic materials with negative
parameters have the oppositely directed phase vector and the  Poynting
vector. By this reason, such a medium is sometimes referred as {\it
backward-wave} (BW) medium. L.I.~Mandelshtam first pointed out to
unusual reflection and refraction laws at interfaces between BW
and conventional media \cite{Mand}, which was recently
observed experimentally \cite{exp2,exp1}. V.G.~Veselago performed an 
electrodynamical study of such a medium \cite{Veselago}, referring
to it as {\it ``left-handed medium"} and proposed, in particular, a
planar slab made of this material as a focusing lens. The concept
of ``perfect" lens from plane plate of BW material has been
developed by \cite{Pendry}.  Ziolkowski and
Heyman \cite{Ziolk}  simulated pulse propagation through a slab
of BW medium, using the FDTD method and re-considered  possibilities  to design
the ``perfect" lens.

Design of media, where the phase and energy velocities point to
the opposite directions has a long history. Actually, this
property exists in slow-wave structures for electronic generators
with extended interaction, backward-wave tubes \cite{Hutter}, as well as in travelling wave antennas \cite{Walter}. Any relations between the phase and group velocities
directions, including  the opposite, can be observed in
two-dimensional periodic structures \cite{Silin}, which in the modern literature are referred as  {\it
photonic crystals} \cite{PBG}.

However, all of these structures exhibit negative dispersion in
such a spectral range, where the wavelength is comparable with the
structure period, and it is possible to consider  only their
effective negative refractive index, expressing that in terms of the slow-wave
factor. The realization of a composite material, where the
structural sizes are much smaller than the wavelength, and an
experimental verification of its properties was described
in \cite{exp2,exp1}.
These new metamaterials, in a certain frequency range, can be considered as
homogeneous media described by some negative permeability and permittivity parameters.
A possibility for a realization of wideband composites with active
inclusions was theoretically considered in \cite{Tret}.

N. Engheta introduced an idea to make a compact cavity resonator
composed of two layers, so that one of them is a usual
 material, and the other one is a BW medium \cite{Engh}.
  If this structure is inserted between two electric walls,
the resonant frequencies of the cavity do not depend on
the total thickness of the two-layered structure, but only on the
 ratio of the tangents of the thicknesses of the
separate layers. Such a property  suggests a possibility to
realize very thin (or thick, if desired) resonators. Obviously, such a
two-layered structure considered as a waveguide would exhibit some
properties which are not met in  waveguides composed of usual
materials. In  a conference presentation \cite{Engh-boulder}
it was  pointed to some peculiarities of this waveguiding
structure, in particular to the fact that in the limit of
thin slabs the propagation factor approximately cancels out from the dispersion
relation.

In this paper we present our results of a detailed study of wave propagation in  two-layered closed
waveguides whose one layer is a usual forward-wave (FW) material and the
other one is a BW (or negative) material. In paper \cite{Smith}
it was  pointed out that a passive BW medium should be dispersive and must
satisfy constrains \cite{Landau}
\begin{equation}
\label{a0}
\begin{array}{ll}
 \frac{d[\varepsilon(\omega)\omega]}{d\omega}>1,\quad &
 \frac{d[\mu(\omega)\omega]}{d\omega}>1.
\end{array}
\end{equation}
However, the spectral properties of the waveguide composed of
FW-BW media were found so unusual in comparison with the  case
of conventional materials, that we have decided first to
restrict ourselves to the  simplest model of
frequency-independent BW medium parameters
 in order to clarify the  role of
relations between geometrical and material parameters of the two media layers.
It appears to be quite acceptable
physically because a small dispersion of $\varepsilon$ and $\mu$ is enough
to satisfy inequalities (\ref{a0}).  Also, these limitations can be
overcome using metamaterials with active inclusions \cite{Tret}.

Probably the most important property of such a waveguide is the existence of
eigenwaves whose slow-wave factor is not
restricted by $\sqrt{\varepsilon\mu}$ as in  usual waveguides, and
whose fields decay exponentially from the media interface within
both of FW and BW layers. The nature of these waves is similar to
surface waves at an interface of vacuum and an isotropic plasma within the spectral range where its
permittivity is negative. Negative permeability $\mu$ makes
possible existence of surface wave of the other, TE polarization. However, our analysis is not restricted to super-slow waves. We
show how the relations between the layer thicknesses and
material parameters influence the waveguide propagation
characteristics of various modes.

\section{EIGENWAVES IN TWO-LAYERED WAVEGUIDE}

\subsection{General relations}

\begin{figure}[ht]
\centering \epsfig{file=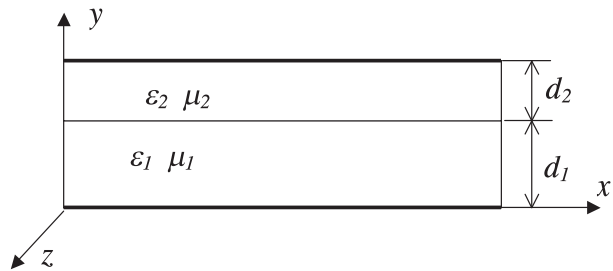, width=8cm}
\caption{Geometry of the problem: a planar waveguide filled by 
two slabs of different materials, one of them is a metamaterial with 
negative parameters.}
\end{figure}

In this paper we consider a plane two-layered waveguide, infinite along  $z$ and
$x$ directions and bounded by electric walls in $z-y$ plane at
distances $d_1$ and $d_2$ from the media interface (see Fig.~1).
The media are characterized by relative permittivities $\varepsilon_1$,
$\varepsilon_2$ and permeabilities $\mu_1$, $\mu_2$.  We will
discuss eigenwaves propagating in $z$ direction whose field
varies depends on  time and the longitudinal coordinate as $\exp(\omega t -k_z z)$.

 Let us first recall the main properties of the modes propagating in the usual
two-layered waveguide. Assuming that $\partial/\partial x=0$,
these waves can be separated into two classes, TE modes, whose
electric field has no longitudinal and perpendicular to the
interfaces components, $E_z=E_y=0$, and TM modes with $H_z=H_y=0$.

\begin{figure}
\centering \epsfig{file=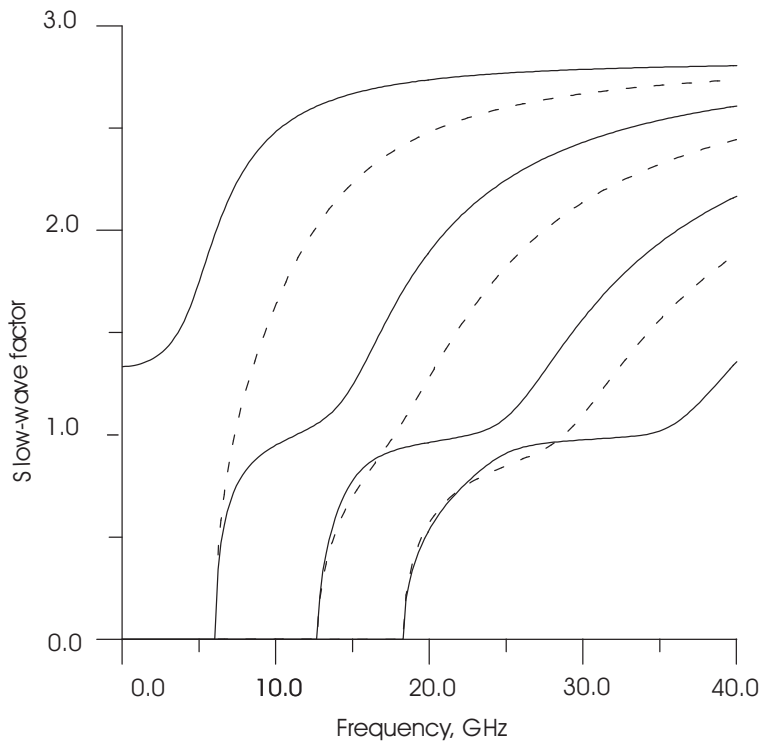, width=10cm}
\caption{Dispersion of TM modes (solid lines) and TE
modes (dashed lines) in a two-layered waveguide, composed of usual
(FW) materials, calculated with
$\mu_1=2$, $\mu_2=1$, $\varepsilon_1=4$, $\varepsilon_2=1$, $d_1=0.5$
cm, $d_2=1$ cm.}
\end{figure}

TE modes satisfy the dispersion relation
\begin{equation}
\label{a2} \frac{\mu_1}{k_{y1}}\tan{k_{y1}d_1}+
\frac{\mu_2}{k_{y2}}\tan{k_{y2}d_2}=0,
\end{equation}
where $k_{yi}=\sqrt{k^2\varepsilon_i\mu_i-k_z^2}\;(i=1,2)$, $k$ is the
wavevector in vacuum.
TM modes are governed by the dispersion relation
\begin{equation}
\label{a1} \frac{k_{y1}}{\varepsilon_1}\tan{k_{y1}d_1}+
\frac{k_{y2}}{\varepsilon_2}\tan{k_{y2}d_2}=0.
\end{equation}
The dominant TM$_0$ mode has no cutoff and its slow-wave factor
has the low-frequency limit
\begin{equation}
\label{a3} n_{TM_0}\to
\sqrt{\frac{d_1\mu_1+d_2\mu_2}{d_1/\varepsilon_1+d_2/\varepsilon_2}},
\end{equation}
 which can be obtained by expanding the tangent functions in (\ref{a1}) in
 Taylor series or by averaging the
permittivities and permeabilities  within the first and second
layers. Other TM and TE modes appear in pairs with the same cutoff
frequencies given by relation
\begin{equation}
\label{a4} \sqrt{\frac{\mu_1}{\varepsilon_1}}\tan{k\sqrt{\varepsilon_1\mu_1}d_1}+
\sqrt{\frac{\mu_2}{\varepsilon_2}}\tan{k\sqrt{\varepsilon_2\mu_2}d_2}=0,
\end{equation}
which is obtained from (\ref{a2}) or (\ref{a1}) under condition
$k_z=0$.
 It is important, that the dispersion properties of the
waveguide modes are determined by resonance phenomena: cutoffs
correspond to different standing-wave resonances of the cross-section
$d_1+d_2$. Let us assume, that the slow-wave factor in unbounded
first-layer medium is larger than in the second layer,
$\varepsilon_1\mu_1>\varepsilon_2\mu_2$.
 When the frequency is only slightly over the cutoff frequency,
 the wave is fast, but
soon its slow-wave factor becomes close to $\sqrt{\varepsilon_2\mu_2}$. The
field is concentrated within the layer with the smaller value of
$\varepsilon\mu$ and the field distribution is described by trigonometric
functions inside both layers. Further increase of the frequency causes a
re-distribution of the field, so that the slow-wave factor tends to
$\sqrt{\varepsilon_1\mu_1}$, see Fig.~2. The field distribution within
layer 2 is now described by exponential functions. However, equations
for TM modes (\ref{a1}) and for TE modes (\ref{a2})
have wave solution only when at least one of the
tangent arguments in Eqs.~(\ref{a2},\ref{a1},\ref{a4}) is real,
$k_{y1}^2>0$ in our case. It is very important in our contents that
in the guide filled with conventional media
the slow-wave factor never exceeds the largest value of
the two refractive indices $\sqrt{\varepsilon_{1,2}\mu_{1,2}}$.

\subsection{Backward-wave layer case: general properties}

What happens, if one of the layers is a BW material? Let us assume that
$\varepsilon_1<0,\, \mu_1<0$. The dispersion relations for TE and TM
modes become
\begin{equation}
\label{a6} \frac{|\mu_1|}{k_{y1}}\tan{k_{y1}d_1}-
\frac{\mu_2}{k_{y2}}\tan{k_{y2}d_2}=0
\end{equation}
and
\begin{equation}
\label{a5} \frac{k_{y1}}{|\varepsilon_1|}\tan{k_{y1}d_1}-
\frac{k_{y2}}{\varepsilon_2}\tan{k_{y2}d_2}=0.
\end{equation}

The same minus sign appears in the cutoff relation (\ref{a4}), as it 
follows from (\ref{a6},\ref{a5}) at $k_z=0$. Now real solutions of
Eqs.~(\ref{a2},\ref{a1},\ref{a4}) are permitted with both
$k_{y1},\, k_{y2}$ being purely imaginary numbers. This means that
the surface waves, whose fields decay exponentially from the
interface between FW and BW layers can propagate in such a
waveguide and there are no upper restrictions for their
propagation constants.

In this respect, let us consider again Eq.~(\ref{a3}), the
low-frequency limit of the slow-wave factor for the fundamental mode.
If the media parameters are all positive, this value is always within the limits
\begin{equation}
\label{between}
\sqrt{\varepsilon_{2}\mu_{2}}\leq\sqrt{\frac{d_1\mu_1+d_2\mu_2}{d_1/\varepsilon_1+d_2/\varepsilon_2}}\leq
\sqrt{\varepsilon_{1}\mu_{1}}.
\end{equation}
However, if we allow negative values of the material parameters, there are
no limits at all:
\begin{equation}
0\leq\sqrt{\frac{d_1\mu_1+d_2\mu_2}{d_1/\varepsilon_1+d_2/\varepsilon_2}}\leq
\infty.
\end{equation}
Very peculiar situations take place in the limiting cases. If
\begin{equation}
\label{crazy} d_1/\varepsilon_1+d_2/\varepsilon_2\rightarrow 0,
\end{equation}
we observe that the capacitance per
unit length of our transmission line (we now consider the quasi-static limit)
tends to infinity. This means that although the voltage drop between the plates
tends to zero, the charge density on the plates remains finite.
This can be understood from a simple observation that if we fix the charge densities
(positive on one plate and negative on the other), the displacement vector is
fixed and, in the quasi-static limit, it is constant across the cross section.
However, the electric field vector is oppositely directed in the
two slabs, if one of the permittivities is negative. In the limiting case
(\ref{crazy}) the total voltage tends to zero. Similarly, in the limiting case
$d_1\mu_1+d_2\mu_2\rightarrow 0$, the inductance per unit lengths tends to zero.

Another interesting observation concerns the case when both
layer thicknesses tend to infinity, that is, the case of waves travelling
along a planar interface between two media.
The dispersion equations reduce to
\begin{equation}
\label{interface} \frac{|\mu_1|}{k_{y1}}-
\frac{\mu_2}{k_{y2}}=0, \quad \mbox{TE modes}, \qquad
\frac{k_{y1}}{|\varepsilon_1|}-
\frac{k_{y2}}{\varepsilon_2}=0, \quad \mbox{TM modes}
\end{equation}
It is well known (and obvious from the above relations) that
surface waves at an interface can exist only if at least one of
the media parameters is negative, an obvious example is an
interface with a free-electron plasma region. If both parameters
are negative, both TE and TM surface waves can exist. A very
special situation realizes if the parameters of the two media
differ only by sign, that is, if $\varepsilon_1=-\varepsilon_2$
and $\mu_1=-\mu_2$. In this case the propagation factor cancels
out from the dispersion relations, because $k_1=k_2$. This means
that waves with any {\it arbitrary} value of the propagation
constant are all eigenwaves of the system at the frequency where
this special relation between the media parameters is realized. A
similar observation was made in \cite{Engh-boulder} as an
approximation in case of small heights $d_{1,2}$. For a media
interface this result is exact.

\begin{figure}
\centering \epsfig{file=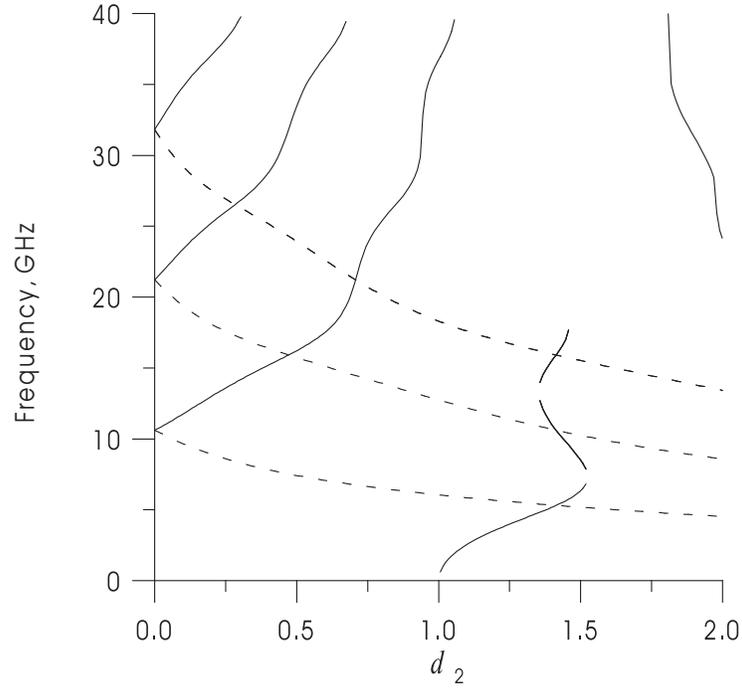, width=10cm}
\caption{Cutoff frequencies versus $d_2$ calculated for
fixed $d_1=0.5$ cm, when the first layer is a FW medium (dashed
line) and BW medium (solid line). The material parameters of the layers
are $\mu_1=\pm 2;\;\mu_2=1;\;\varepsilon_1=\pm
4;\;\varepsilon_2=1$.}
\end{figure}

Next, let us study how the cutoff frequencies depend on the layer thickness.
Let us fix the thickness of the first, BW layer, and consider dependence
of the cutoff frequencies on the thickness of the second layer.
The results are shown in Fig.~3. The cutoff frequencies $F_c$ for the
usual two-layer
waveguide are presented also for comparison (dashed curves). When
$d_2=0$, the cutoff frequencies are equal for the waveguides
filled with BW and FW materials. An increase of $d_2$ causes
a decrease of the effective thickness of the two-layered waveguide if
the first layer is a BW medium. This  leads to an increase of $F_c$. In other words,
we can say that the cutoff corresponds to a resonance condition for
waves travelling in the vertical direction, along $y$ axis. Since the phase velocity is directed oppositely in the two layers, the phase shift is partially
compensated, hence the electric thickness gets smaller and the cutoff frequency
increases (see solid curves in Fig.~3). However, further increase of $d_2$
leads to a compensation of this negative negative contribution to the phase shift,  and $F_c$ becomes again smaller.

\section{TE modes}

\begin{figure}
\centering \epsfig{file=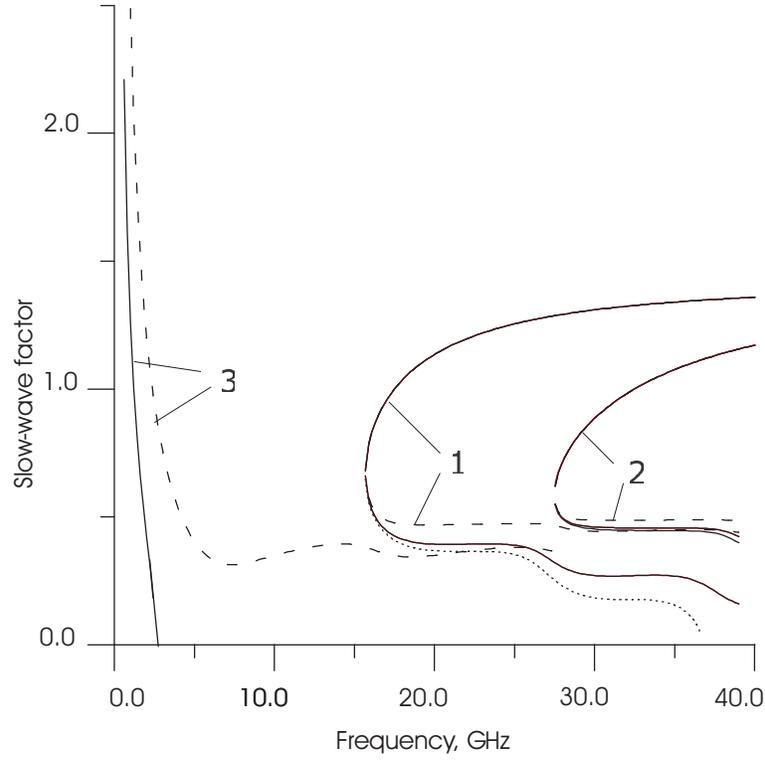, width=10cm}
\caption{Dispersion of TE modes, when the first layer is
a BW material and the second layer is a usual material:
$\mu_1=-2;\;\mu_2=1;\;\varepsilon_1=-4;\;\varepsilon_2=1$, $d_1=0.5$
cm. The thicknesses of the second layer are: $d_2=1$ cm (dotted
line), $d_2=1.1$ cm (solid line), $d_2=2$ cm (dashed line).
 }
\end{figure}

Let us consider dispersion of TE modes (see Fig.~4), calculated
with different thicknesses of the second (FW) layer. One notable
difference from the usual two-layered waveguide is a change of
dispersion sign (see curves 1,2). It is caused by the opposite
directions of the longitudinal components of the energy transport
within FW and BW layers.
 The frequency point where  dispersion changes sign,
corresponds to the situation when the total energy flows are equal
within the first and second layers.  The upper parts of the
dispersion characteristics are nearly the same for waveguides with
different $d_2$, because in this case the field is concentrated
mainly  within the first layer. However, the lower  parts depend
on $d_2$, see solid, dotted and dashed lines in Fig.~4.

Another important new feature, already noted above,  is a
possibility of propagation of  waves whose slow-wave factor
exceeds $\sqrt{\varepsilon_1\mu_1}$ (we assume, as before, that
$\varepsilon_1\mu_1>\varepsilon_2\mu_2$). Dispersion
characteristics of such a super-slow wave for different $d_2$ are
shown by curves~3, Fig.~4. We observe that the slow-wave factor
tends to infinity if $k\to 0$. When its propagation is possible?
Assuming $k\to 0,\, k_z\neq 0$, Eq.~(\ref{a6}) becomes
\begin{equation}
\label{d1}
 F(k_z)={|\mu_1|\tanh{k_zd_1}-\mu_2\tanh{k_zd_2}\over{k_z}}=0.
\end{equation}
Consider two limiting cases, when $k_zd_{1,2}$ are small or large.
In the limit $k_zd_{\rm min}\to \infty$ ($d_{\rm
min}=\min({d_1,d_2}))$, $F(k_z)>0$ due to the assumption
$|\mu_1|>\mu_2$.
 For the second limiting case
there  may be two possibilities, when $k_z=0,\;k\neq 0$ and
$k_z/k\to n_0\neq 0,\;k\to 0$. If $k_z=0,\;k\neq 0$
 we can expand the
tangents in Taylor series and obtain
 $\tan{k_{y1}d_1}\simeq  k\sqrt{\varepsilon_1\mu_1}d_1$,
  $\tan{k_{y2}d_2}\simeq  k\sqrt{\varepsilon_2\mu_2}d_2$,
  $F(k_z)\approx |\mu_1|  d_1-\mu_2 d_2$.
 Obviously, $F(k_z)$  can be negative  for small $k_z$ only if the condition
\begin{equation}
\label{d2}
 |\mu_1|d_1 < \mu_2d_2,\qquad  (\mbox{for} \quad |\mu_1|> \mu_2)
\end{equation}
 is satisfied. In this case function $F(k_z)$ changes sign and there is
 a solution of (\ref{a6}).
  Next let $k_z/k$ be  equal to some nonzero value $n_0$ at
  $k\to 0$. Then $k_{y1}=k\sqrt{\varepsilon_1\mu_1-n_0^2}\to 0$,
   $k_{y2}=k\sqrt{\varepsilon_2\mu_2-n_0^2}\to 0$, and we come again to
    relation (\ref{d2}).

 Numerical solution of equation $F(k_z)=0$
confirms existence of nonzero roots $k_z$, corresponding to
very large slow-wave factors if condition (\ref{d2}) is satisfied.
  In our example, just when  $\mu_2 d_2$ becomes larger
than $\mu_1 d_1=1$ cm (we have taken $\mu_2 d_2=1.1$ cm), a new mode
appears (see curve 3, solid line). Further increase of the
thickness $d_2$ causes a shift of the dispersion curve to higher
frequencies (dashed curve).

Still another  peculiarity of the spectrum of two-layered
waveguides filled with FW-BW materials is the existence of a
non-dispersive wave (recall our assumption that both of FW and BW
materials are non-dispersive).
To study this possibility, let us first note that non-dispersive solutions are only
possible if the arguments of the two tangent functions in (\ref{a6})
are equal (so that there is no dependence on the  wavenumber $k$).
If this condition is satisfied, the tangent functions can be cancelled, and
the eigenvalue equation (\ref{a6}) can be easily solved. The result
for the slow-wave factor reads
\begin{equation}
\label{e1}
 n_c=\sqrt{\frac{|\mu_1|\mu_2(|\mu_1|\varepsilon_2-\mu_2|
\varepsilon_1|)}{\mu_1^2-\mu_2^2}}.
\end{equation}
Next, let us check if the arguments of the tangent functions can be
indeed equal for  this solution. Substituting  (\ref{e1}),
the arguments of the tangent functions in (\ref{a6}) read
\begin{equation}
\sqrt{{\varepsilon_1\mu_1-\varepsilon_2\mu_2\over{\mu_1^2-\mu_2^2}}}
|\mu_1|d_1\quad\mbox{and}\quad
\sqrt{{\varepsilon_1\mu_1-\varepsilon_2\mu_2\over{\mu_1^2-\mu_2^2}}}\mu_2d_2.
\end{equation}
These values must be equal for a non-dispersive solution.
We conclude that the non-dispersive wave with the propagation factor
given by (\ref{e1}) exists
if the following two conditions are satisfied:
\begin{equation}
\label{e2}
\begin{array}{l}
|\mu_1|d_1=\mu_2d_2 \\
(|\mu_1|\varepsilon_2-\mu_2|\varepsilon_1|)(\mu_1^2-\mu_2^2)> 0.
  \end{array}
\end{equation}
The first condition guaranties the existence of a solution, and if the
second condition  is satisfied, the solution is a real number corresponding
to a propagating mode.
It can be seen also that such a solution describes a surface wave
with an exponential field distribution in both of the media if
\begin{equation}
\label{e3} (\mu_1^2-\mu_2^2)(\varepsilon_1\mu_1-\varepsilon_1\mu_1)<0.
\end{equation}
In other cases the field distribution is described by
trigonometric functions.

 \begin{figure}
\centering \epsfig{file=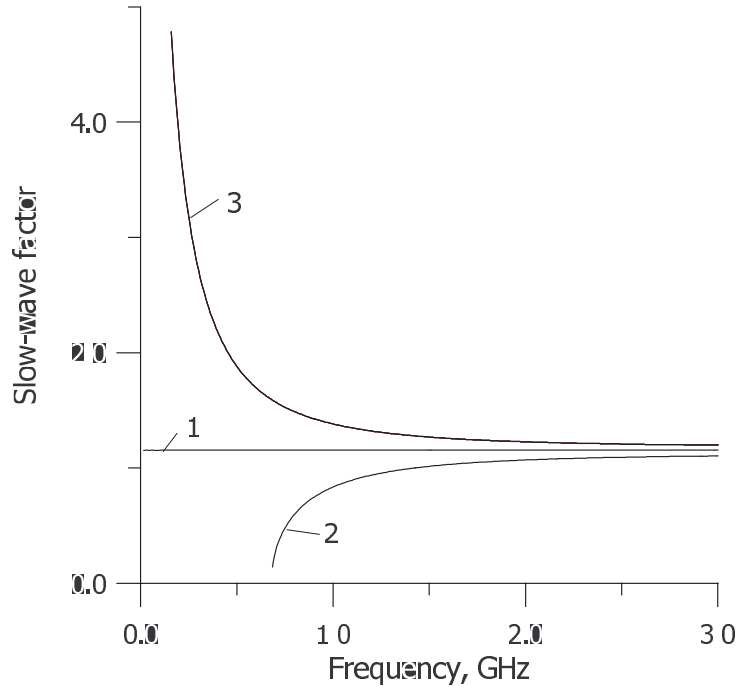, width=10cm}
\caption{Dispersion characteristics of TE modes,
calculated as
$\mu_1=-2;\;\mu_2=1;\;\varepsilon_1=-4;\;\varepsilon_2=3$,
$d_1=0.1$~cm, $d_2$=0.2~cm or $d_1=1$~cm, $d_2$=2~cm (curve~1),
$d_1=1$~cm, $d_2$=1.95~cm (curve~2) and $d_1=1$~cm,
$d_2=2.05$~cm (curve~3).
 }
\end{figure}

Let us discuss the properties of this non-dispersive wave. Its
propagation constant does not depend on the frequency, which means
that  such a wave has no cutoff. Furthermore, its existence is
not connected with the total thickness $d_1+d_2$, but only with
the relation $d_2/d_1=|\mu_1|/\mu_2$. It is illustrated by curve~1,
Fig.~5, calculated at $d_1=0.1$ cm. No other waves propagate
within the spectral range presented in Fig.~5.
 This wave disappears under any small deviation of
either $d_2$ or $d_2$ violating  relation $|\mu_1|d_1=\mu_2d_2$, but
it is not so sensitive to  the values of $\varepsilon_1$ and $\varepsilon_2$, it is
enough that inequality (\ref{e3}) is satisfied.

Consider next what happens under a small deviation of the second
layer thickness, violating  relation $|\mu_1|d_1=\mu_2d_2$. If, as
above, $d_1=0.1$ cm, $d_2=0.2$ cm, taking $d_2$ smaller than 0.2
cm we find that the non-dispersive wave disappears, and no other
waves propagate in the considered spectral range. Taking $d_2$
larger than 0.2 cm, the non-dispersive wave disappears also, but
a new super-slow wave appears, whose slow-wave factor
is larger than presented in Fig.~5. Let us consider thicker layers
with $d_1$=1~cm, $d_2$=2~cm. Obviously, the slow-wave factor of
the  non-dispersive wave remains the same. Let the thicknesses of
the first layers be 1~cm and that of the second be 1.95~cm. In
this case the wave has a low-frequency cutoff, determined by
Eq.~(\ref{a4}), and its dispersion characteristic tends to the
dispersion curve of the non-dispersive wave (see curve~2, Fig.~5).
Taking  $d_2$=2.05~cm, we observe a super-slow wave at low
frequencies, whose dispersion also tends to the case of the
non-dispersive wave at large frequencies (see curve~3). Thus,
small deviations of $d_2$ dramatically change the  dispersion
properties of the eigenwaves at low frequencies.

\section{TM modes}

\begin{figure}
\centering \epsfig{file=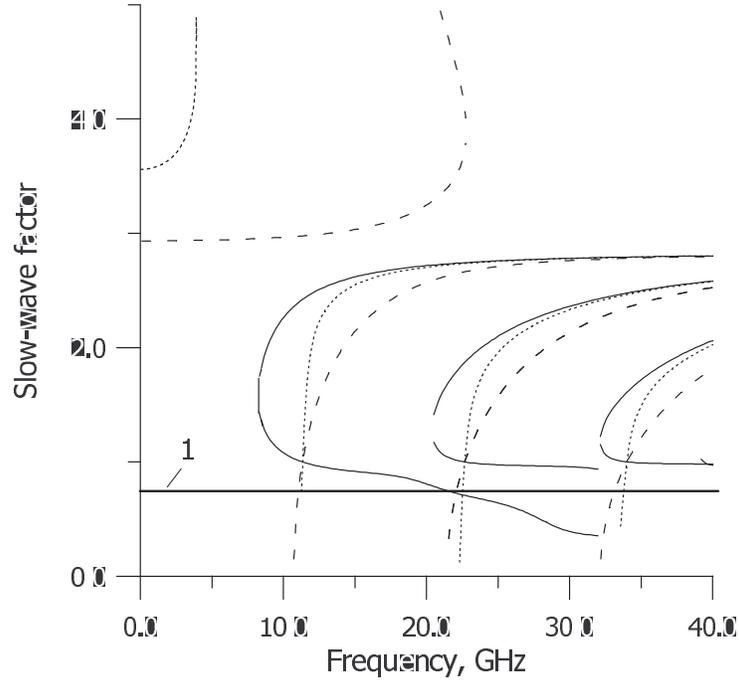, width=10cm}
\caption{Dispersion of TM modes, when the first layer is
a BW material and the second layer is a usual material.
 The thicknesses of the second layer are: $d_2=1$ cm (solid
line), $d_2=0.01$ cm (dashed line), $d_2=0.05$ cm (dotted line).
The other parameters of the waveguide are the same as in Fig.~4. 
The thick curve~1 shows the non-dispersive TM mode.
 }
\end{figure}

Dispersion characteristics of TM modes are presented in Fig.~6. As
for TE modes, there exist modes  whose slow-wave factor is
restricted by
$n_{\max}=\max{\left(\sqrt{\varepsilon_1\mu_1},\sqrt{\varepsilon_2\mu_2}\right)}$.
They may change the sign of  dispersion (solid curves, Fig.~6) or
not  (dashed and dotted lines) depending of the value of $d_2$ (we
assume that $d_1=0.5$~cm  is fixed). The change of the dispersion
sign is caused by the opposite directions of the energy transport
within the FW and BW layers, as was discussed above for the TE modes.


Analysis of Eq.~(\ref{a5}) shows, that, similarly to the case of
TE modes, a non-dispersive TM mode exists under conditions
\begin{equation}
\label{f1}
\begin{array}{ll}
|\varepsilon_1|d_1=\varepsilon_2d_2,\\
(|\varepsilon_1|\mu_2-\varepsilon_2|\mu_1|)(\varepsilon_1^2-\varepsilon_2^2)>0.\\
 \end{array}
\end{equation}
Its slow-wave factor is constant and equals to
\begin{equation}
\label{f2}
 n_c=\sqrt{\frac{|\varepsilon_1|\varepsilon_2
(|\varepsilon_1|\mu_2-\varepsilon_2|\mu_1|)}{\varepsilon_1^2-\varepsilon_2^2}}.
\end{equation}
 This wave is localized near the interface of the two media slabs
 if
\begin{equation}
\label{f3} (|\varepsilon_1|-\varepsilon_2)(|\varepsilon_1|\mu_1-\varepsilon_2\mu_2)<0.
\end{equation}
In other cases the field distribution is described by
trigonometric functions. As a non-dispersive TE mode, such a TM
mode can exist even if the thicknesses of the layers are much smaller
than the wavelength. Its dispersion is shown by  curve~1,
calculated for the same $\varepsilon$ and $\mu$ as the other curves in
Fig.~6.


\begin{figure}
\centering \epsfig{file=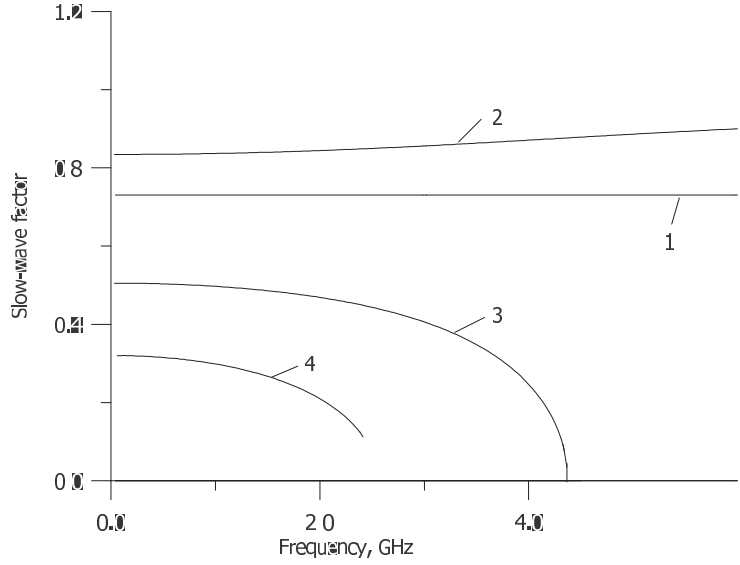, width=10cm}
\caption{Dispersion of the dominant TM$_0$ mode,
calculated at the same $\varepsilon_{1,2},\, \mu_{1,2}$, and $d_1$ as in Fig.~6 and
$d_2=2$ cm (curve~1), $d_2=3$ cm (curve~2), $d_2$=1.3 (curve~3),
$d_2=1.1$ cm (curve~4).
 }
\end{figure}

Next we consider the properties of  the dominant TM$_0$ mode, if
one of the layers is a BW material. The quasi-static limit
(\ref{a3}) becomes
\begin{equation}
\label{g1} n_{TM_0}\to
\sqrt{\frac{d_1|\mu_1|-d_2\mu_2}{d_1/|\varepsilon_1|-d_2/\varepsilon_2}}.
\end{equation}
However, now such a wave can propagate only under condition
\begin{equation}
\label{g2}
(d_1|\mu_1|-d_2\mu_2)(d_1/|\varepsilon_1|-d_2/\varepsilon_2)>0,
\end{equation}
when the square root is real. As already noted above, in contrast to  the ordinary
waveguide, where
\begin{equation}
\label{g3}
 \sqrt{\varepsilon_2\mu_2}<n_{TM_0}<\sqrt{\varepsilon_1\mu_1}
\end{equation}
 (under assumption
$\sqrt{\varepsilon_2\mu_2}<\sqrt{\varepsilon_1\mu_1}$), the value
$n_{TM_0}$ given by (\ref{g2}) is not restricted by  relation
(\ref{g3}). Let us fix $d_1$ and consider how the dispersion of
this wave changes in dependence of $d_2$. If
$d_2>d_1|\varepsilon_1|/\varepsilon_2=2$~cm, the slow-wave factor
increases with $d_2$ and tends to 1 if $d_2\to\infty$ at small
frequencies. It is illustrated by Fig.~7, where  curve 1,
calculated at $d_2$=2 cm, corresponds to the non-dispersive wave,
and curve~2 corresponds to a larger value of the second layer
thickness $d_2$=3 cm.

The slow-wave factor decreases with $d_2$ while condition
(\ref{g2}) is satisfied, that is $d_2>d_1|\mu_1|/\mu_2=1$ cm, but
$d_2<d_1|\varepsilon_1|/\varepsilon_2=2$~cm for chosen parameters.
In this case  a high-frequency cutoff is observed, see curves~3,4
in Fig.~7.

Thus, an important difference between non-dispersive TM and TE
modes is that the TE mode disappears at small frequencies if the
relation $ |\mu_1|d_1=\mu_2d_2$ is violated (see Fig.~5). In
contrast, the non-dispersive TM mode under violation of the
condition $ |\varepsilon_1|d_2=\varepsilon_2d_1$ continuously
changes its dispersion and exists at $k\to 0$.

\begin{figure}
\centering \epsfig{file=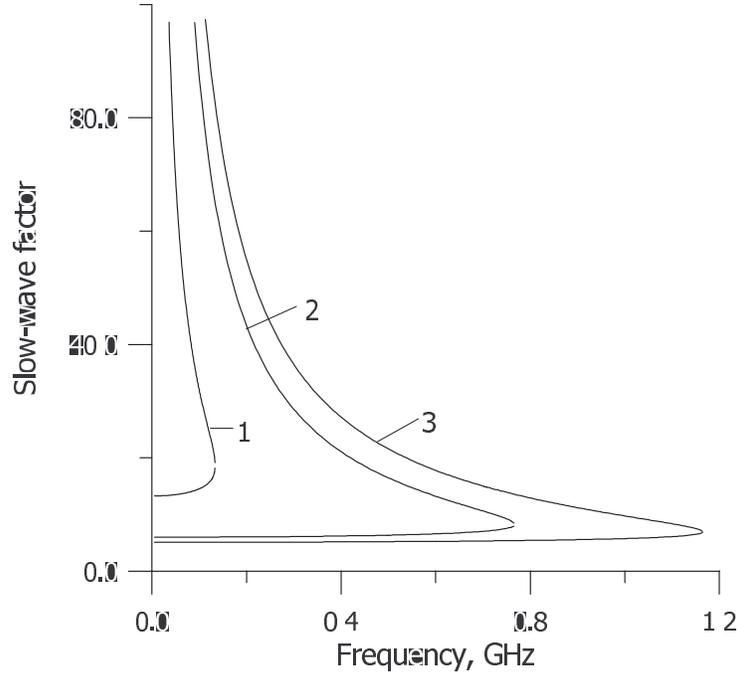, width=10cm}
\caption{Dispersion of the super-slow TM mode, calculated at
$d_2=0.12$ cm (curve~1), $d_2=0.1$ cm (curve~2), and $d_2=0.09$ cm
(curve~3).
 }
\end{figure}

Another possibility of the existence of a wave without a low-frequency
cutoff, following from (\ref{g2}), is the condition
$d_2<d_1\varepsilon_2/|\varepsilon_1|$. This case is especially important, because
there propagation of a super-slow mode is permitted, like for the TE
mode. Let us show, that existence of such a wave becomes indeed possible.
Consider the function which determines the left-hand side of
Eq.~(\ref{a5}):
\begin{equation}
\label{g4}
 F(k_z)=\frac{k_{y1}}{|\varepsilon_1|}\tan{k_{y1}d_1}-
\frac{k_{y2}}{\varepsilon_2}\tan{k_{y2}d_2}.
\end{equation}
In the limiting case  $k_zd_{\min}\gg 1\; (d_{\min}=\min({d_1,d_2}))$,
$F(k_z)<0$ because $|\varepsilon_1|>\varepsilon_2$. Since there is a solution $k_z=k
n_{TM_0}$ at $k\to 0$, we can expand the tangents in Taylor series
at small $k$: $\tan{x}\sim x+x^3/3+...$. Taking into account that
\begin{equation}
 \label{g5}
\begin{array}{l}
 \varepsilon_1\mu_1-n_{TM_0}=
 \frac{|\varepsilon_1|
 d_2(\varepsilon_1\mu_1-\varepsilon_2\mu_2)}{d_2|\varepsilon_1|-d_1\varepsilon_2},\\
 \varepsilon_2\mu_2-n_{TM_0}=
 \frac{\varepsilon_2 d_1(\varepsilon_1\mu_1-\varepsilon_2\mu_2)}{d_2|\varepsilon_1|-d_1\varepsilon_2},
 \end{array}
\end{equation}
 and $\tan{k_{y1}d_1}\approx k\sqrt{\varepsilon_1\mu_1-(n_{TM_0})^2}d_1$,
 $\tan{k_{y2}d_2}\approx k\sqrt{\varepsilon_2\mu_2-(n_{TM_0})^2}d_2$,
  we find that the first term of the expansion gives zero value of
  $F(k_z)$, and taking the second term we obtain
\begin{equation}
\label{g6}
 F(k_z)\approx
\frac13\frac{|\varepsilon_1|\mu_1-\varepsilon_2\mu_2}{d_2|\varepsilon_1|-d_1\varepsilon_2}d_1^2d_2^2
\left(|\varepsilon_1|d_1-\varepsilon_2d_2\right).
\end{equation}
Since we have assumed $|\varepsilon_1|>\varepsilon_2$, the condition  $F(k_z)>0$ at small
$k_zd_{\min}$
becomes
\begin{equation}
(|\varepsilon_1|d_2-\varepsilon_2d_1)(|\varepsilon_1|d_1-\varepsilon_2d_2)>0.
\end{equation}
This can be realized if both expressions in the brackets are of the 
same sign. 
Numerical calculations confirm existence of a wave, whose
slow-wave factor dramatically increases with $k\to 0$ only when 
\begin{equation}
\label{g7}
\begin{array}{l}
|\varepsilon_1|d_2-\varepsilon_2d_1>0, \\
|\varepsilon_1|d_1-\varepsilon_2d_2>0,
 \end{array}
 \end{equation}
which in our case corresponds to $d_2<0.125$. The
super-slow wave is seen at Fig.~6, dashed and dotted curves.
 Fig.~8 illustrates dispersion properties of super-slow
TM-modes, calculated for different $d_2$ at small frequencies.
 The smaller is $d_2$, the larger frequency band where such a mode is observed.
Numerical analysis has shown the absence of super-slow solutions
for the other pair of conditions, giving $F(k_z)>0$, namely,
\begin{equation}
\label{g8}
\begin{array}{l}
|\varepsilon_1|d_2-\varepsilon_2d_1<0 ,\\
|\varepsilon_1|d_1-\varepsilon_2d_2<0.
 \end{array}
 \end{equation}
In this case wave solutions also exist, but they are normal waves with 
moderate propagation constants and weak dispersion (at low frequencies).

\section{Conclusion}

We have considered the eigenmodes in a layered waveguide
containing a layer of a backward-wave metamaterial, which has negative and
real material parameters.
We have found important
differences between the eigenmode spectra in ordinary and FW-BW
two-layered waveguide, which can be summarized as following:

\begin{enumerate}
\item
In a FW-BW waveguide both  TE and TM modes can change the
dispersion sign. This is  possible because the energy transport directions
are  opposite in FW and BW layers, so there exists a spectral point, where the
power flows in the two layers compensate each other.
\item
Under certain relations between the permeabilities and thicknesses
of FW and BW layers there exists a non-dispersive TE mode without a
low-frequency cutoff. Its slow-wave factor is constant and does
not depend on the layers thicknesses.
\item
In contrast with the ordinary two-layered waveguide, where always
exists a dominant TM$_0$ mode without a low-frequency cutoff, in the FW-BW
waveguide its analog disappears under certain conditions. In
addition, such a wave can be non-dispersive under certain
relations between the permittivities and thicknesses of FW and BW
layers.
Furthermore, this mode has a high-frequency cutoff under certain conditions.

\item
There exist both TE and TM super-slow waves, whose slow-wave
factor is not restricted by the values of the permittivities
and permeabilities of the layers. The fields of these waves decay
exponentially in both  FW and BW layers from their interface
 in case of large propagation constants.
 It is remarkable, that
such super-slow modes are caused not by large values of the
permeability or permittivity, like it takes place near a resonance in
ferrite or plasma, but by the layer thickness effects.

The super-slow TE mode has a high-frequency cutoff at $k_z$=0, or
its slow-wave factor tends to a limit, determined by
the non-dispersive TE mode.

The slow-wave factor of the super-slow TM mode has a bottom restriction,
determined by the dominant TM$_0$ mode.

\end{enumerate}

\bibliography{ntw}

\end{document}